\documentstyle[twocolumn,aps]{revtex}
\input psfig.sty
\def\={&=&}

\def\ltsim{\vbox {\hbox{\lower 0.8\baselineskip \hbox{$<$}} \break
		 \hbox{\lower 0.2\baselineskip \hbox{$\sim$}} } }
\def\gtsim{\vbox {\hbox{\lower 0.8\baselineskip \hbox{$>$}} \break
                 \hbox{\lower 0.2\baselineskip \hbox{$\sim$}} } }
\def\k{{\bf k}}

\def\q{{\bf q}}

\def\r{{\bf r}}

\def\vs{{\bf v}_s}
\def\v{{\bf v}}

\def\A{{\bf A}}
\def\H{{\bf H}}
\def\K{{\cal K}}
\begin{document}
\draft

\twocolumn[\hsize\textwidth\columnwidth\hsize\csname %
@twocolumnfalse\endcsname

\title{
Is the nonlinear Meissner effect unobservable?}
\author{M. -R. Li,$^{a,b}$ P. J. Hirschfeld$^{a,c}$ and P. W\"olfle$^a$}

\address{
$^a$Institut f\"ur Theorie der Kondensierten Materie, Universit\"at
Karlsruhe, 76128 Karlsruhe, Germany\\
$^b$Department of Physics, Nanjing University, Nanjing 210093, P.R.China\\
$^c$Department of Physics, University of Florida, Gainesville, FL 32611, 
USA \\}

\maketitle
\begin{abstract}
We examine the effects of nonlocal electrodynamics for a $d$-wave 
superconductor on the field dependence of the magnetic penetration depth. 
The linear field dependence predicted in the local limit, commonly known 
as the nonlinear Meissner effect, is instead found to be quadratic, 
$\delta\lambda\sim H^2$ for fields below a  crossover scale $H^*$.
This crossover is shown to be geometry dependent, and for most orientations
of the screening currents,  is of the same order as or greater than $H_{c1}$, 
implying that the nonlinear Meissner effect can not be observed. 
For special orientations where the current flows along the nodal 
directions, however, the nonlinear Meissner effect may be recovered. 
\end{abstract}
]
Measurements of the linear term in the temperature dependence of the 
electromagnetic penetration depth \cite{Hardy} $\delta \lambda(T)\equiv 
\lambda(T)-\lambda(0)\sim T$ played a pivotal role in the identification of 
the $d$-wave symmetry of the cuprate superconductors.  It was always clear, 
however, that they were sensitive only to the depletion of the superfluid 
fraction by the thermal excitations of quasiparticles with momenta near the 
order parameter nodes, and hence provided information only on the energy gap 
on these regions of the Fermi surface.  Furthermore, in a  $d$-wave 
superconductor with tetragonal crystal symmetry, the penetration depth is 
independent of the angle $\theta$  the supercurrent makes with the crystal 
axes since the linear response tensor is isotropic; the gap {\it shape} is 
thus not directly reflected in the angle dependence of the effect.
%\vskip .2cm

Yip and Sauls\cite{YipSauls} argued, however, that information about the gap
shape in an unconventional superconductor with line nodes could be obtained 
by measuring the nonlinear (magnetic field-dependent) penetration depth 
$\delta \lambda(H)=\lambda(H)-\lambda(0)$, and predicted an unusual field 
dependence,
\begin{eqnarray}
{\delta \lambda (H)\over \lambda (0)}\simeq {\zeta_\theta\over \sqrt{2}}
{H\over H_0}, \label{ys}
\end{eqnarray}
where $\zeta_\theta$ is a (weak) angle-dependent function, $H$ is the applied
magnetic field, and $H_0=3\Phi_0/(\pi^2 \xi_0\lambda_0)$ is of the order the 
thermodynamic critical field of the system ($\Phi_0$ is the flux quantum,
$\xi_0$ the coherence length and $\lambda_0$ the penetration depth).
Later, it was shown\cite{Xhuetal} that impurities alter this prediction, 
leading to an asymptotic $\delta \lambda \sim H^2$ dependence, but only at 
concentrations such that the linear term in the $H\rightarrow 0$ temperature 
dependence $\delta\lambda(T)\sim T$ is replaced by $\delta\lambda(T)\sim
T^2$ \cite{Grossetal,Prohammer,felds}.
%\vskip .2cm

From an experimental standpoint, the search for confirmation of this 
fundamental manifestation of $d$-wave symmetry has been controversial.   
In 1995, Maeda et al.  reported the first observation of a linear $H$ term 
on $Bi_2Sr_2CaCu_2O_8$ $(BSCCO-2212)$ films \cite{Maedaetal}, but the results 
were questionable since they were obtained at relatively high temperatures, 
where the Yip-Sauls  theory predicts a $\delta\lambda\sim H^2$. The clean 
$d$-wave nonlinear Meissner effect should be obtained, according to the 
simple theory, only for temperatures below a crossover $T\ll E_{nonlin}$, 
with $E_{nonlin}=v_sk_F\simeq (H/H_0)\Delta_0$ and $\Delta_0$ the gap maximum, 
giving  typically $E_{nonlin}\simeq 1K$ for the cuprates, for fields close 
to $H_{c1}$, the lower critical field. Measurement of a linear-$H$ term was 
also claimed at much higher temperatures  for $YBa_2Cu_3O_{7-\delta}$ ($YBCO$) 
samples \cite{Maeda2}. It is important to recall that several extrinsic effects 
can give rise to spurious field dependences, most importantly contributions 
from trapped vortices which enter near sample edges. An extrinsic contribution 
$\delta\lambda\sim\sqrt{H}$ to the effective penetration depth can arise from 
coupling of the rf field used in resonant coil experiments to the pinned 
vortices \cite{Campbellpendepth}. Such dependences were also found in 
measurements at much lower temperatures and on higher quality single crystals 
of $YBCO$ by Carrington and Giannetta \cite{CarringtonGiannetta}. Finally, 
a measurement to detect higher-harmonic nonlinear modes in the ac response 
predicted on the basis of Yip-Sauls theory by $\breve{\rm Z}$uti\'{c} and Valls 
\cite{ZuticValls}, apparently less sensitive to the effects of trapped flux, 
failed to observe the predicted signal \cite{Goldmanprivate}. 
%\vskip .2cm
 
We therefore take the point of view that the available experimental data do 
not support the existence of a nonlinear Meissner effect in the best cuprate 
samples, and ask why. Our answer is that the effects of nonlocal electrodynamics, 
neglected in the work of Yip and Sauls, serve to cut off the (local) nonlinear 
Meissner effect as the field is lowered, replacing the linear-$H$ response with 
a quadratic one below a  crossover field $H^*$.  In most experimental geometries, 
this crossover is of the order of the lower critical field $H_{c1}$, 
 meaning the 
$\delta\lambda(H)\sim H$ behavior cannot generically be observed in the 
Meissner state.  However, we show that nonlocal effects are negligible in 
geometries with currents parallel to a nodal direction.
%\vskip .2cm

{\it Nonlocal electrodynamics.}  The Yip-Sauls description of the nonlinear 
Meissner effect is performed in the local limit, i.e. the $d$-wave Cooper 
pairs, although extended objects, are assumed to respond to the electromagnetic 
field at a single point at the center of the pair.   Such an approximation 
is normally justified in classic type-II superconductors, where $\lambda_0$ is 
much larger than $\xi_0$. Although the cuprates are strongly type-II 
superconductors, with small in-plane $\xi_0\simeq{\cal O}(15A)$ and large 
in-plane $\lambda_0\simeq {\cal O}(1500A)$, effects of nonlocal electrodynamics 
are to be seen in several special situations, as a result of the gap nodes, 
along which the ``effective coherence length" $v_F/\pi\Delta_k$ diverges 
\cite{KL}.  Such effects were shown to significantly suppress the shielding 
supercurrents in measurements of the ac Meissner effect \cite{PHW} in 
unconventional superconductors with line nodes, with application to heavy 
fermion materials. In the context of the cuprates, it was shown by Zucarro 
et al.\cite{Scharnberg} that significant variations in the electromagnetic 
response of clean $d$-wave superconductors were to be expected when nonlocal 
effects were included. But it was Kosztin and Leggett\cite{KL} who first 
pointed out that the very signature of a superconductor with line nodes, the 
linear-$T$ term in the dc penetration depth, was modified at the lowest 
temperatures \cite{SD}. The crossover temperature below which this occurs 
is $E_{nonloc}\simeq \kappa^{-1}\Delta_0$, where $\kappa\equiv 
\lambda_0/\xi_0$ is the Ginzburg-Landau parameter. The characteristic 
crossover field defined by $E_{nonloc}=E_{nonlin}$ is given by $H^*\simeq 
\kappa^{-1} H_0\simeq H_{c1}$. One might therefore expect that nonlocal 
effects ``cut off" the nonlinear Meissner effect. This observation needs, 
however, to be supported by a full calculation, which we now sketch.

%\vskip .2cm
{\it General current response.}  The current response for a $d$-wave 
superconductor may be derived from a full calculation of the BCS free energy 
for fixed external field $\H$ corresponding to vector potential $\A$. We 
have performed such a calculation (including both nonlocal and nonlinear 
effects analytically) under the assumptions of i) negligible spatial 
variations of the amplitude of the order parameter, $\Delta_\k(\r)$; and 
ii) slow spatial variations of the superfluid velocity $\vs$. The details 
of the derivation will be given elsewhere \cite{LHWPRB}, but the final 
result for the Fourier component of the current in an infinite system at 
{\it constant} ${\vs}$ is ${\bf j}(\q)=-\K(\q,\vs,T) \A(\q)$, where
\begin{eqnarray}
{\K(\q,\vs,T)\over c/(4\pi \lambda_0^2)}=1+{2T\over nm}
\sum_{\omega_n,{\bf k}}{\bf k}^2_{\parallel}\frac{z_\k^2+\xi_+\xi_- +
\Delta_+\Delta_-}{[z_\k^2-E^2_+][z_\k^2-E^2_-]} \, ,
\label{bubble}
\end{eqnarray}  
with $\lambda_0=\sqrt{ mc^2/4\pi ne^2}$ the {\it local} penetration depth 
at $T=0$ in {\it linear} response theory, $n$ the electron density, $m$ the 
electron mass, $c$ the speed of light, $\k_\parallel$ the projection of 
$\k$ onto the surface, $z_\k\equiv i\omega_n+\vs\cdot\k_F$, 
$E_\pm^2\equiv \xi_\pm^2+\Delta_\pm^2$, $\xi_\pm\equiv\xi_{\k\pm\q/2}$, 
and $\xi_\k={k^2\over 2m}-\mu$ ($\mu$ the chemical potential).  This is 
the standard linear response result with all quasiparticle Matsubara 
energies $\omega_n$ modified by the semiclassical Doppler shift 
$\vs\cdot\k_F$ \cite{MakiTsuneto}. For the real physical system in the 
Meissner state, a surface boundary is present and $v_s$ decays with the 
distance from the surface $y$. In the linear, local approximation, 
$v_s(y)=(e\lambda_0 H/mc) \exp(-y/\lambda_0)$ decays exponentially and 
$v_s(y=0)$ is proportional to the external field, neither of which hold 
generally. But we may imagine the system to be subdivided into many layers, 
in each of which $v_s$ is roughly constant. Furthermore, we assume 
{\it specular} scattering surface boundary condition on the quasiparticle 
wavefunctions. In this case, the response is given by Eq. (\ref{bubble}). 
The geometry which we study first  is such that the magnetic field is along 
the $c$ axis and thus $\vs$ lies in the $ab$ plane; the normal of the 
boundary plane $\hat q$ is also in the $ab$ plane, forming an angle 
$\theta$ with the $b$ axis. 
We will discuss the situation where {\bf $H$} is perpendicular to the 
$c$ axis later. 
%\vskip .2cm

To a first approximation, we neglect the space dependence of $v_s$ and 
replace it by its value at the surface $v_s\simeq v_s(y=0)=e\lambda_0 H/mc$ 
as given by the solution to the linear, local electrodynamics problem. Now 
we separate out the $T=0$ local, linear response as $\K(\q,\vs,T)=
c/(4\pi\lambda_0^2)+\delta\K(\q,\vs,T)$, and then {\it define} the nonlinear, 
nonlocal penetration depth to be
\begin{eqnarray}
\lambda_{\rm spec}=\int^\infty_0{H(y)\over H(0)}dy\simeq\lambda_0
-{8\over c}\int^\infty_0 dq\frac{\delta{\cal K}(\q,{\bf v}_s,T)}
{(\frac{1}{\lambda^2_0}+q^2)^2}.
\label{pene1}
\end{eqnarray} 
 This expression has the virtue of reducing exactly 
to the nonlocal expression of Kosztin and Leggett\cite{KL} if the 
$\vs$-dependence is neglected, and (qualitatively) to the nonlinear 
expression of Yip and Sauls \cite{YipSauls} if the $q$-dependence is 
neglected.  It is worthwhile reviewing the latter case.  When $T\ll v_sk_F$ 
we obtain
\begin{eqnarray}
{\delta{\cal K}(\q\rightarrow 0,{\bf v}_s,T)\over c/(4\pi\lambda_0^2)}=
-\frac{\zeta_\theta}{\sqrt{2}}\frac{v_sk_F}{\Delta_0} \, ,
\label{response9}
\end{eqnarray}  
where $\zeta_\theta={1\over 2}\sum_{l=\pm 1} |\cos\theta+l\sin\theta|^3$. 
Thus, 
\begin{eqnarray}
\frac{\delta\lambda^{({\rm loc})}_{\rm spec}}{\lambda_0}
=\frac{\lambda^{({\rm loc})}_{\rm spec}-\lambda_0}{\lambda_0}
\simeq\frac{\zeta_\theta}{2\sqrt{2}}\frac{v_sk_F}{\Delta_0}
\simeq\frac{3}{2}\frac{\zeta_\theta}{\sqrt{2}}\frac{H}{H_0}\; .
\label{penetra2}
\end{eqnarray}  
We recall that Yip and Sauls \cite{YipSauls} and Xu et al. \cite{Xhuetal} 
defined penetration depth from its initial decay: 
$\lambda=-H(0)/[dH(y)/dy]|_{y=0}$ which is equivalent to the definition in 
Eq. (\ref{pene1}) only in the linear limit. Using their definition we find 
that the prefactor 3/2 in Eq. (\ref{penetra2}) is modified to 3. At this 
stage, we may improve our approximation by restoring the spatially varying 
nature of $v_s$. We follow Yip et al. to solve directly a nonlinear London 
equation 
\begin{eqnarray}
\nabla^2 \v_s={4\pi\over c}\K(0^+, \v_s,T)\v_s=
{\v_s\over \lambda^2_0}(1-\frac{\zeta_\theta}{\sqrt{2}}
\frac{k_F}{\Delta_0}v_s) \, ,
\label{london1}
\end{eqnarray}  
under the boundary condition $dv_s/dy|_{y=0}=(e/mc)H$. Using the relation
$dv_s/dy=(e/mc)H(y)$ and their definition for the penetration depth again, 
we see that $\delta\lambda/\lambda_0=2\frac{\zeta_\theta}{\sqrt{2}}
\frac{H}{H_0} $. So in the local limit we get qualitatively the same 
nonlinear correction to the penetration depth as Yip and Sauls \cite{YipSauls}, 
comparing to Eq. (\ref{ys}), but with a larger prefactor. This deviation 
arises from the perturbative treatment of the response of the superconductor 
to the $\A(\q\neq 0)$ modes in the present theory. The nonlinear effects 
obtained at $T\ll v_sk_F$, due to the Doppler shift coming from the coupling 
of the $\A(\q=0)$ mode to the quasiparticles, are nonperturbative results. 
A discussion of this point can be found in Ref. \cite{LHWPRB}.

%\vskip 0.2cm
{\it Low-energy scaling properties.}  
Explicit analytical results for the full response in the approximation of
constant $v_s$ may now be obtained by expanding $\delta\K$ for low energies, 
$v_s k_F, q v_F, T\ll\Delta_0$. A 2-parameter scaling relation may be derived, 
\begin{eqnarray}
\delta{\cal K}(q,v_s,T)=-\frac{c}{8\pi\lambda^2_0}\frac{T}{\Delta_0}
\sum_{l=\pm 1} u^2_{\theta l}
F_\lambda({\alpha w_{\theta l}\over T},{\varepsilon u_{\theta l}\over T}),
\nonumber\\
F_\lambda(z_1,z_2)=\frac{\pi}{4}z_1+[\ln (e^{z_2}+1)+\ln (e^{-z_2}+1)]\nonumber\\
-\int^{z_1}_{0}dx[f(x-z_2)+f(x+z_2)][1-(\frac{x}{z_1})^2]^{1/2},
 \label{scaling}
\end{eqnarray}  
with $\alpha=v_Fq/(2\sqrt{2})$, $\varepsilon=v_sk_F/\sqrt{2}$,
$u_{\theta l}=|\cos\theta+l \sin\theta|$,
$w_{\theta l}=|\sin\theta-l \cos\theta|$ and $f(x)=1/(1+e^x)$.
 We consider primarily the limit $T\ll \alpha,\varepsilon$.
It is straightforward to see that for $z_2\gg 1$  
\begin{eqnarray}
F_\lambda(z_1,z_2)\simeq
\left \{
\begin{array}{ll} 
z_2\;\;\; &  z_1\ll z_2 \\
{\pi z_1\over 4}+z_2(z^2_2+\pi^2)/(6z^2_1) \;\;\; &  z_2\ll z_1\, .
\end{array}
\right.
\label{scaling1}
\end{eqnarray} 
The scaling form (\ref{scaling}) may now be inserted into (\ref{pene1}) to find
the penetration depth. Its field dependence turns out to be determined by the 
ratio $\alpha w_{\theta l}/\varepsilon u_{\theta l}$, with a typical $q$ set 
to $1/\lambda_0$ in (\ref{pene1}) by the denominator.
If we define two new dimensionless parameters $h_{\theta l}=(u_{\theta l}
/w_{\theta l}) h$, $l=\pm 1$,  with $h\equiv mv_s\lambda_0\rightarrow 
(3/\pi)\kappa H/H_0$, we get $\delta\lambda_{\rm spec}=\sum_{l=\pm 1}
\delta\lambda^{(l)}_{\rm spec}$, where
 \begin{eqnarray}
\frac{\delta\lambda^{(l)}_{\rm spec}}{\lambda_0}\simeq
{\pi\over 4\sqrt{2}}u^3_{\theta l}\kappa^{-1} h+c_{\theta l1}
&\kappa^{-1}&\frac{1}{h^2}
+c_{\theta l2}\kappa{1\over h^4}{T^2\over \Delta_0^2} \nonumber \\
&&{\rm for}\;\;\; h_{\theta l} \gg 1, \label{lamnonlin}
\end{eqnarray} 
with $c_{\theta l1}=0.008 w^3_{\theta l}$ and $c_{\theta l2}=0.006 
w^3_{\theta l}/u^2_{\theta l}$, and 
\begin{eqnarray}
\frac{\delta\lambda^{(l)}_{\rm spec}}{\lambda_0}\simeq
{\pi\over 8\sqrt{2}}d_{\theta l1}\kappa^{-1}
+d_{\theta l2}&\kappa^{-1}& h^2 + 
d_{\theta l3} \kappa {T^2\over\Delta_0^2} \nonumber \\
&&{\rm for}\;\;\; h_{\theta l} \ll 1, \label{lamnonloc}
\end{eqnarray} 
with $d_{\theta l1}=0.5 u^2_{\theta l}w_{\theta l}$, $d_{\theta l2}=1.1 
u^4_{\theta l}/w_{\theta l}$ and $d_{\theta l3}=0.47 u^2_{\theta l}/w_{\theta l}$. 
The singularity which leads to the linear term in Eq. (\ref{lamnonlin}) 
if nonlocal effects are simply neglected is seen to be cut off by the constant 
term in Eq. (\ref{lamnonloc}). This happens at $h_{\theta l}\sim 1$. For 
$\theta$ not too close to a nodal value, $w_{\theta l}$ and $u_{\theta l}$ 
are of order unity for both $l=\pm 1$, and the crossover field 
may be simply determined from $h=1$, $H^*\sim\kappa^{-1}H_0$.   
The thermodynamical critical field $H_c\simeq H_0$ is the geometric mean 
$\sqrt{H_{c1}H_{c2}}$ of the upper critical field $H_{c2}=\Phi_0/(2\pi\xi_0^2)$ 
and the lower critical field $H_{c1}=\Phi_0/(2\pi\lambda_0^2)$.  Thus the 
crossover field {\it for such geometries} is of the same order as the lower 
critical field, $H^*\simeq H_{c1}$.  Since the Meissner state is unstable to 
the Abrikosov vortex state for fields above $H_{c1}$, the linear-$H$ term in 
$\delta\lambda$ is asymptotically unobservable in these geometries. In the 
Meissner state it is, in fact, replaced by a quadratic variation, as seen 
in (\ref{lamnonloc}). In Fig. 1, we have plotted the penetration depth 
calculated in Kosztin and Leggett's geometry $\theta=0$, according to the exact 
scaling function  given in (\ref{scaling}), to explicitly exhibit this effect.

When $\theta$ is near any node, however, $u_{\theta l}$ can be much larger than
$w_{\theta l}$ and the nonlocal energy scale  much smaller than the nonlinear
one, leading to a very small crossover field $H^*\simeq H_{c1}w_{\theta l}
/u_{\theta l}\ll H_{c1}$. We conclude that the nonlinear Meissner effect may
be observable in such special geometries, e.g. a sample with a dominant (110) 
surface and $H\parallel c$.
\begin{figure}[h]
\begin{picture}(200,180)
\leavevmode\centering\includegraphics{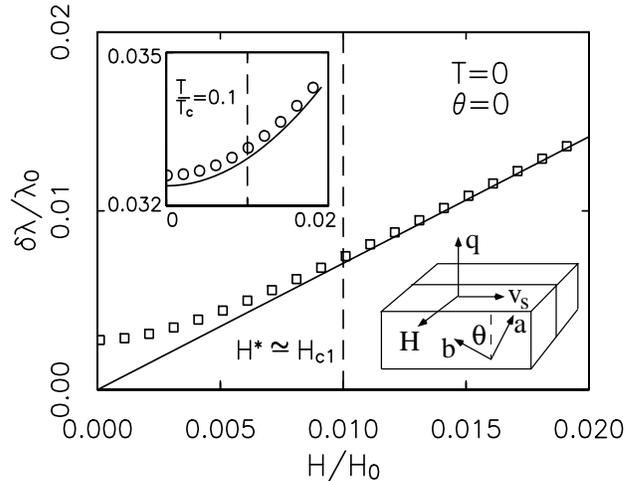}
\end{picture}
\caption{  
Normalized penetration depth correction $\delta\lambda/\lambda_0$ vs. normalized
magnetic field $H/H_0$ at $T=0$ for a sample with $\bf H$ parallel to a $(010)$ 
surface ($\theta=0$) and Ginzburg-Landau parameter $\kappa=10^2$, hence lower 
critical field $H_{c1}\simeq 10^{-2}H_0$.  Squares: full response; solid line: 
nonlocal effects neglected.  Insert: $T=0.1T_c$. Circles: full response; solid 
line: nonlocal effects neglected. Dashed line: $H^*\simeq H_{c1}$.}
\end{figure} 
%\vskip .2cm
There is another special situation in the quasi-2D cuprates in which the
characteristic energy of nonlocal effects becomes very small and the previous
discussion might not be expected to apply.  If we assume that the quasiparticles 
are strictly confined to the $ab$ plane and if the experimental configuration 
is dominated by a (001) surface with $\H\parallel ab$, then $E_{nonloc}^{(ab)}=
\q\cdot\v_F=0.$  In this case, however, the  magnetic field is not screened 
($\lambda\rightarrow\infty$) \cite{GK}. Whenever  the quasiparticles are
allowed to move along the $c$-direction we expect  nonzero $E_{nonloc}^{(ab)}$ 
\cite{schopohlcomm}. This can be demonstrated directly for any model of  
coherent transport along the $c$-axis. 
The materials of greatest current interest are the cuprate materials  
$YBCO$ and $BSCCO-2212$.   In the $YBCO$ case, it is reasonable to treat the 
system as weakly-3D.  The characteristic nonlocal energy is then $v_{Fc}q\simeq
\xi_{0c}\Delta_0/\lambda_0$  much smaller than in the (010) case since the 
$c$-axis coherence length is $\xi_{0c}\simeq 3A$ as opposed to the in-plane 
coherence length of $\xi_{0}\simeq 15A$. On the other hand, the lower critical 
field is also much smaller for this geometry, $H_{c1}^{(ab)}\simeq \Phi_0/(2\pi 
\lambda_0\lambda_{0c})$. Using $\lambda_{0c}~\simeq ~0.5-0.8\times10^4 A$, 
we again find a large crossover field $H^{*(ab)}\simeq (\xi_{0c}/\xi_0) H^*
\simeq 0.5-1 H_{c1}^{(ab)}$, making it still 
impossible from a practical point of view to extract a linear-$H$ term
\cite{Cooper}. 

%\vskip .2cm
The BSCCO system is so anisotropic that it is questionable whether the above argument
applies. A full treatment of this problem awaits a generally accepted theory of 
the (incoherent)
$\hat c$-axis transport in the normal state.
%\vskip .2cm

In conclusion, with the exception of the claim by Maeda et al. 
\cite{Maedaetal,Maeda2}, all 
attempts to measure the nonlinear Meissner effect, a fundamental consequence 
of $d$-wave symmetry, have failed. As the order parameter symmetry is by now 
well-established in these materials, an explanation is required. To this end, 
we have considered the interplay of nonlinear and nonlocal effects in the 
electromagnetic response of a $d$-wave superconductor in the Meissner regime.
This situation differs from the similar problem considered by Amin et al.
\cite{Aminetal} in the vortex phase due to the fact that the nonlinear and 
nonlocal energies $E_{nonlin}$ and $E_{nonloc}$ are not independent in their 
case, as a consequence of the fluxoid quantization condition. By contrast, 
in the Meissner state, while the nonlocal scale is set by intrinsic normal-state 
parameters, i.e. $T_c$ and $\kappa$, the nonlinear energy $E_{nonlin}\simeq 
v_sk_F$ is effectively tunable by an external field.  We have shown that the 
nonlocal terms in the response generically cut off the nodal singularity in 
the nonlinear response, eliminating the linear-$H$ term in the penetration 
depth.

%\vskip .2cm
Neither the current theory nor the Yip-Sauls result, both based on a consideration 
of the shielding currents in the Meissner phase, can explain the results of Maeda 
et al.  Maeda's results  were obtained on samples which did not clearly exhibit a
$\delta\lambda(T;H\rightarrow 0)\sim T$ behavior in any range, and therefore should 
not show nonlinear effects even according to the Yip - Sauls theory.  Furthermore, 
they were performed at temperatures of 9K and above, and a linear-$H$ behavior was 
reported up to 50K. As we show in Fig. 1, even for temperatures as low as $0.1T_c$, 
no linear term {\it should} be visible even if nonlocal effects are neglected.  
Therefore we must conclude that Maeda et al. observed an extrinsic effect, probably 
related to trapped vortices. Very recently,  new high-resolution resonant coil
measurements\cite{CarringtonGiannetta,BHBL} on very clean $YBCO$  samples
observed (in some geometries) linear $H$ terms with magnitudes close to the 
Yip-Sauls prediction \cite{CarringtonGiannetta,BHBL}, but simultaneously measured 
qualitatively inconsistent temperature dependence.  These authors also observed 
effects attributed to trapped vortices \cite{CarringtonGiannetta}, and we 
therefore conclude that the linear term must be of extrinsic origin. Clearly, 
a test of the nonlinear effects which is independent of trapped flux is desirable.
The harmonics of the nonlinear response predicted by $\breve{\rm Z}$uti\'{c} and 
Valls in principle provide such a test, but to date,  a high resolution nonlinear
transverse magnetization experiment designed to observe these resonances has 
failed to do so \cite{Goldmanprivate}. This null result is consistent with the 
current theory, but by itself cannot confirm it due to uncertainties regarding 
the identification of $H_{c1}$ \cite{Goldmanprivate}.  

Finally, in all experiments up to now the orientation of the supercurrent was 
along the crystal axes. Our analysis suggests the best possibility to observe 
the effect is in geometries with $v_s$ along the nodal directions.

%\vskip .2cm
 The authors are grateful to M. Franz, W. Hardy, and N.
Schopohl for helpful communications, and particularly to R. Giannetta and 
A. Goldman for discussion of their unpublished results.  Partial support was 
provided by NSF-DMR-96--00105 and the A. v. Humboldt Foundation.

\end{document}